\documentclass[referee]{aa}
\usepackage{graphicx}
\usepackage{txfonts}

\def\lsim{\lower.5ex\hbox{$\; \buildrel < \over \sim \;$}}
\def\gsim{\lower.5ex\hbox{$\; \buildrel > \over \sim \;$}}

\begin{document}
\title{Evolution of the quasi-periodic oscillation frequency in GRO J1655-40 -- Implications for accretion disk dynamics }
\author{Sandip K. Chakrabarti\inst{1,2}, Dipak Debnath\inst{2}, Anuj Nandi\inst{2,3} and P.S. Pal \inst{2}}
\offprints{Prof. Sandip K. Chakrabarti}
\institute{S. N. Bose National Centre for Basic Sciences, Salt Lake, Kolkata 700098, India
\and
Indian Centre for Space Physics, Chalantika 43, Garia Station Rd., Kolkata, 700084, India
\and
On leave from Indian Space Research Organization, Bangalore
   \date{Received ; accepted }
\abstract
{Low and intermediate frequency quasi-periodic oscillations (QPOs) are thought to be due to 
oscillations of Comptonizing regions or hot blobs embedded in Keplerian disks. Any movement of these
perturbations is expected  systematically to change the QPO frequency.}
{Our goal is to find systems where such a systematic drifts have been observed. We also
try to find the real cause of such drifts and whether they shed some light on the 
accretion disk dynamics.}
{Using archival data of the recent outburst of GRO J1655-40, we report the
presence of such systematic drifts not only during the rising phase from 
the $25^{th}$ of February 2005 to the $12^{th}$ March 2005, when the QPO frequency
monotonically increased from $82$mHz to $17.78$Hz but also in the decline phase from
the $15^{th}$ September 2005 to the $5^{th}$ of October 2005, when the QPO frequency decreased
from $13.14$Hz to $34$mHz.}
{We fitted the frequency drifts with the propagatory oscillating shock solution.
In the shock-oscillation solution, the frequency is inversely proportional
to the infall time scale from the shock location. We obtained the shock location and strength through 
such a fit. 
}
{The astonishing smoothness of the variation of the QPO frequency
over a period of weeks directly supports the view that it may due to the drift of an oscillating shock
rather than the movements of a blob inside a differentially rotating disk.}
}
\titlerunning{Quasi Periodic Oscillations and Accretion Disk Dynamics}
\authorrunning{Chakrabarti, Debnath, Nandi and Pal}
\maketitle

\section{Introduction}

The galactic nano-quasar GRO J1655-40 is an interesting low mass X-ray binary (LMXB) with a primary mass 
$M = 7.02\pm0.22~M_\odot$ (Orosz \& Bailyn 1997) and the companion star mass 
= $2.3~M_\odot$ located at a distance of $D = 3.2 \pm 0.2$~kpc (Hjellming \& 
Rupen 1995). The disk has an approximate inclination angle of $\theta = 69.5^\circ\pm0.1^\circ$ 
(Orosz \& Bailyn 1997) to the line of sight. 
In the last week of February 2005 it became X-ray active 
(see Shaposhnikov et al. 2007 and references therein) and remained so for 
the next 260 days before returning to the hard state. 
In this Letter, we thoroughly analyse the data of the first two weeks of 
the very initial stage (rising phase) and the last three weeks of the final stage 
(decline phase) of the 2005 outburst. We clearly observe very smooth day to day variation of the
QPO frequency in these two phases. We propose that a satisfactory explanation of this
behavior can be obtained if we assume that an oscillating shock which is sweeping inward
through the disk in the rising phase and outward in declining phase is responsible for the QPO. In the next
Section, we briefly present an overview of QPOs observed in the black hole candidates. 
In \S 3,  we present the shock oscillation solution for the generation of QPOs.
In \S 4, we present the observational results in detail and show our model fit of the
QPO frequencies from day to day. We interpret the results  and extract the shock parameters. 
In \S 5, we give a coherent description of the rising and the declining phases of the outburst.

\section{Low and intermediate frequency QPOs in black hole candidates}

Observations of low and intermediate frequency quasi-periodic oscillations (QPOs) 
in black hole candidates have been reported quite extensively in the literature. One
satisfactory model shows that the oscillation of X-ray intensity is actually due to the
oscillation of the post-shock (Comptonizing) region
(Molteni, Sponholz \& Chakrabarti 1996 [hereafter MSC96]; Chakrabarti \& Manickam 2000 [hereafter CM00]). 
Perturbations inside a Keplerian disk also have been assumed to be the cause of low-frequency QPO also 
(e.g., Trudolyubov, Churazov and Gilfanov 1999; see, Swank 2001 for a review). The numerical simulations
of low-angular momentum accretion flows including the thermal cooling effects (MSC96; Chakrabarti, 
Acharyya \& Molteni 2004, hereafter CAM04) or dynamical cooling (through outflows, e.g., Ryu, 
Chakrabarti \& Molteni 1997) show clearly that the shocks oscillated with frequencies similar to 
the observed QPO frequencies. Not only were the shock locations found to be a function of the cooling 
rate (MSC96), they were found to propagate when viscous effects are turned on (Chakrabarti \& Molteni 1995). 

\section{The properties of low and intermediate frequency QPOs from shock oscillations}

It has been argued in the past that steady, propagating and oscillating shocks can form in a low angular momentum 
flow (e.g., CAM04 and references therein). In the shock oscillation solution (MSC96, CM00; CAM04) of QPOs, 
the oscillations take place at a frequency inverse to the infall time in the post-shock region (i.e.,
the region between the shock at $r=r_s$ and the horizon). In a shock-free low angular momentum flow, 
this infall time is $t_{infall} \sim r_s/v = r_s(r_s-1)^{1/2}$, where $v=1/(r_s-1)^{1/2}$ is the free-fall
velocity in a pseudo-Newtonian potential (Paczynsk\'i \& Wiita, 1980) $\phi_{PN}=-1/(r_s-1)$ Here,
distance, velocity and time are measured in units of the Schwarzschild radius $r_g=2GM/c^2$, the
velocity of light $c$ and $r_g/c$ respectively and where, $G$ and $M$ are the universal constant
and the mass of the black hole. However, in the presence of a significant angular momentum 
capable of producing centrifugal pressure supported shocks around a black hole, the 
velocity is reduced by a factor of $R$, the compression ratio $R=\rho_-/\rho_+$, where, 
$\rho_-$ and $\rho_+$ are the densities in the pre-shock and the post-shock flows,
because of the continuity equation $\rho_- v_-= \rho_+ v_+$ across a thin shock.

In the presence of a shock, the infall time in the post-shock region is therefore given by
$$
t_{infall}\sim  r_s/v_+ \sim  R r_s(r_s-1)^{1/2} 
\eqno{(1)}
$$ 
(CM00; Chakrabarti et al. 2005). Of course, to trigger the oscillation, the accretion rate 
should be such that the cooling time scale roughly match the infall time scale (MSC96).
Thus, the instantaneous QPO frequency $\nu_{QPO}$ (in $s^{-1}$) is expected to be
$$
\nu_{QPO} = \nu_{s0}/t_{infall}= \nu_{s0}/[R r_s (r_s-1)^{1/2}]. 
\eqno{(2)}
$$
Here, $\nu_{s0}= c/r_g=c^3/2GM$ is the inverse of the light crossing time of the black hole 
of mass $M$ in $s^{-1}$ and $c$ is the velocity of light. In a drifting shock scenario, 
$r_s=r_s(t)$ is the time-dependent shock location given by
$$
r_s(t)=r_{s0} \pm v_0 t/r_g.
\eqno{(3)}
$$
Here, $r_{s0}$ is the shock location when $t$ is zero and $v_0$ is the shock velocity 
(in c.g.s. units) in the laboratory frame. The positive sign in the second term is to be used for an outgoing 
shock in the declining phase and the negative sign is to be used for the in-falling shock 
in the rising phase. Here, $t$ is measured in seconds from the first detection of the QPO. 

The physical reason for the oscillation of shocks appears to be
a 'not-so-sharp' resonance between the cooling time scale in the post-shock region and the infall time 
scale (MSC96) or the absence of a steady state solution
(Ryu, Chakrabarti \& Molteni, 1997). In both the cases, the QPO 
frequency directly gives an estimate of the shock location (Eq. 1). 
The observed rise of the QPO frequencies with luminosity (e.g., Shaposhnikov \& Titarchuk 2006, 
hereafter ST06) is explained easily in this model since an enhancement 
of the accretion rate increases the local density and thus the cooling rate. 
The resulting drop of the post-shock pressure reduces the shock location
and increases the oscillation frequency. In CM00 and Rao et al. (2000) it was shown that QPOs 
from the higher energy Comptonized photons, thought to be from
the post-shock region, (Chakrabarti \& Titarchuk, 1995), have a higher $Q$ value. 
The latter model requires two components, one
Keplerian and the other having an angular momentum lower than the Keplerian (referred to  hereafter
as sub-Keplerian), and explains a wide variety of observations of
black hole candidates (Smith, Heindl, \& Swank 2002; Smith,  Heindl, Markwardt \& Swank 2001; 
Smith, Dawson \& Swank 2007). As the shocks are the natural solutions to this sub-Keplerian component,
explanation of QPOs from shocks is justified.
When the Rankine-Hugoniot relation is not exactly satisfied at the shock
or the viscous transport rate of the angular momentum is different on both sides, 
the mean shock location would drift slowly due to a difference in pressure
on both sides. In the rising phase of the outburst, a combination of the ram pressure of the incoming 
flow and rapid cooling in the post-shock region (which lowers the 
thermal pressure) pushes the oscillating shock inward. 
In the decline phase, the Keplerian disk itself recedes outward creating a lower
thermal pressure in the post-shock region. In this case, the shock drifts outward.

\section {Observational results and analysis}

We concentrate on the publicly available data of $51$ observational IDs (corresponding to observations of
a total of $36$ days) of GRO J1655-40 acquired with the RXTE
Proportional Counter Array (PCA; Jahoda et al., 1996). Out of these IDs, $27$ are of the rising phase 
(from MJD 53426 to MJD 53441) and $24$ are of declining phase (from MJD 53628 
to MJD 53648). We extracted the light curves (LC), the PDS and the energy spectra from the
good detector unit PCU2 which also happens to be the best-calibrated. We used the
FTOOLS software package Version 6.1.1 and the XSPEC version 12.3.0. 
For the timing analysis (LC \& PDS), we used the Science Data of the Normal mode 
($B\_8ms\_16A\_0\_35\_H$) and the Event mode ($E\_125us\_64M\_0\_1s$, $E\_62us\_32M\_36\_1s$). 
To extract LC from Event mode data files, we used ``sefilter" task and for the
normal mode data files, we used ``saextrct" task. For the energy spectral analysis, 
the {\bf ``Standard2f"} Science Data of PCA was used. For PCA background estimation 
purpose the ``pcabackest" task was used while to generate the response files 
the ``pcarsp" task was utilized. For the rebinning of the `pha' files 
created by the ``saextrct" task, we used the ``rbnpha" task. For all the analysis,
we kept the hydrogen column density ($N_{H}$) fixed at 7.5$\times$ 10$^{21}$ 
atoms cm$^{-2}$ and the systematics at $0.01$.

\begin {figure}[t]
\vskip 0.8 cm
\centering{
\includegraphics[width=8.0cm]{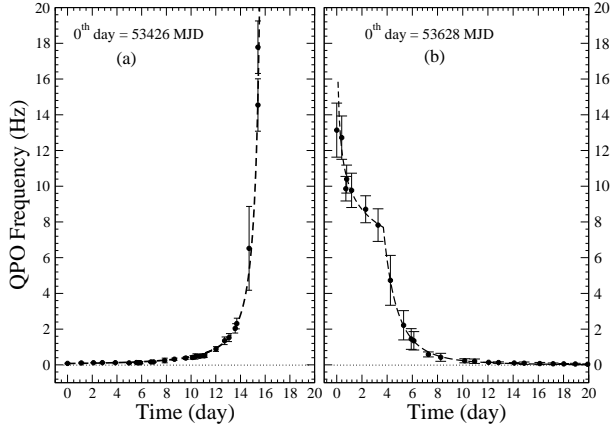}}
\caption{Variation of QPO frequency with time (in day)
(a) of the rising phase and (b) of the declining phase.
Error bars are FWHM of fitted Lorentzian curves in the
power density spectrum. The dotted curves are the solutions from oscillating
and propagating shocks. While in (a) the shock appears to be drifting at a
constant speed towards the black hole, in (b) the shock initially moves very
slowly and then extends at a roughly constant acceleration. According to the
fitted solution, the shock wave goes behind the horizon on the $16.14$th day, about
$15$ hours after the last observed QPO.}
\end {figure}

Figures 1(a-b) show the variation of the QPO frequencies in (a) the rising and (b) the declining phases
of the outburst. The full widths at half maxima of the fitted QPOs have been used as the error bars.
In the rising phase (a), the $0^{th}$ day starts on MJD=53426. The fitted curve represents our fit
with Eqs. (2-3) which requires that the shock is launched at $r_s=1270$ which 
drifts slowly at $v_0=1970$cm s$^{-1}$. On the 
$15^{th}$ day after the outburst starts, the noise was high, but we could clearly observe two
different QPO frequencies with a very short time interval. 
At the time of the last QPO detection ($15.41^{th}$ day) 
at $\nu=17.78$Hz, the shock was found to be located at $r \approx 59$. 
The strength of the shock $R$, which may be strong at the beginning with $R=R_0\sim 4$
should become weaker and ideally $R\sim 1$ at the horizon $r=1$, as it is impossible to maintain
density gradient on the horizon.  If for simplicity we assume the variation of the shock strength 
as $1/R\rightarrow 1/R_0 + \alpha t_d^2$, where $\alpha$ is a very small number limited by the 
time in which the shock disappears (here $t_{ds} \sim 15.5$days).
Thus, the upper limit of $\alpha \sim (1-1/R_0)/t_{ds}^2 =0.75/t_{ds}^2 = 0.003$. 
We find that for a best fit, $\alpha \sim 0.001$ and the reduced $\chi^2=0.96$.  However,
the fit remains generally good ($\chi^2= 1.71$ for $xs0=1245$ and $v0=1960$cm/s) even 
with a shock of constant strength ($R=R_0$). Hence 
for a generally good fit the number free parameters could be assumed to be three: the shock strength, initial shock
location and the shock velocity.

In the declining phase (Fig. 1b), the QPO frequency on the first day
($MJD=53631$) corresponds to launching the shock at $ \sim r_s=40$. It evolves
as $ \nu_{QPO} \sim t_d^{-0.2}$. Since $ \nu_{QPO} \sim r_s^{-2/3}$ (Eq. 1), 
the shock was found to drift very slowly with time ($r_s \sim t_d^{0.13}$) until 
about $t_d=3.5$ day where the shock location was $\sim r=59$. There is a discontinuity in the
behavior at this point whose possible origin is obtained from spectral studies presented below.
After that, it moves out roughly at a constant acceleration ($r_s \sim t_d^{2.3}$) 
and the QPO frequency decreases as $\nu_{QPO}\sim r_s^{-2/3} \sim t_d^{-3.5}$.
Finally, when the QPO was last detected, on $t_d=19.92$th day ($MJD=53648$), 
the shock went as far as $r_s=3100$ and the oscillation could not
 be detected any longer. The strength of the shock was kept at $R=4$. The reduced $\chi^2$ for the
fit is given by $0.236$.
In Figs. 2(a-b), we present the dynamic PDS where the vertical direction indicates the QPO frequency. 
Results of five dwells are given in both the rising and the declining phases. 
The grayscale has been suitably normalized so as to identify the QPO features prominently.

\begin {figure}[h]
\centering{
\vskip -1.0 cm
\includegraphics[width=6.5cm,angle=270]{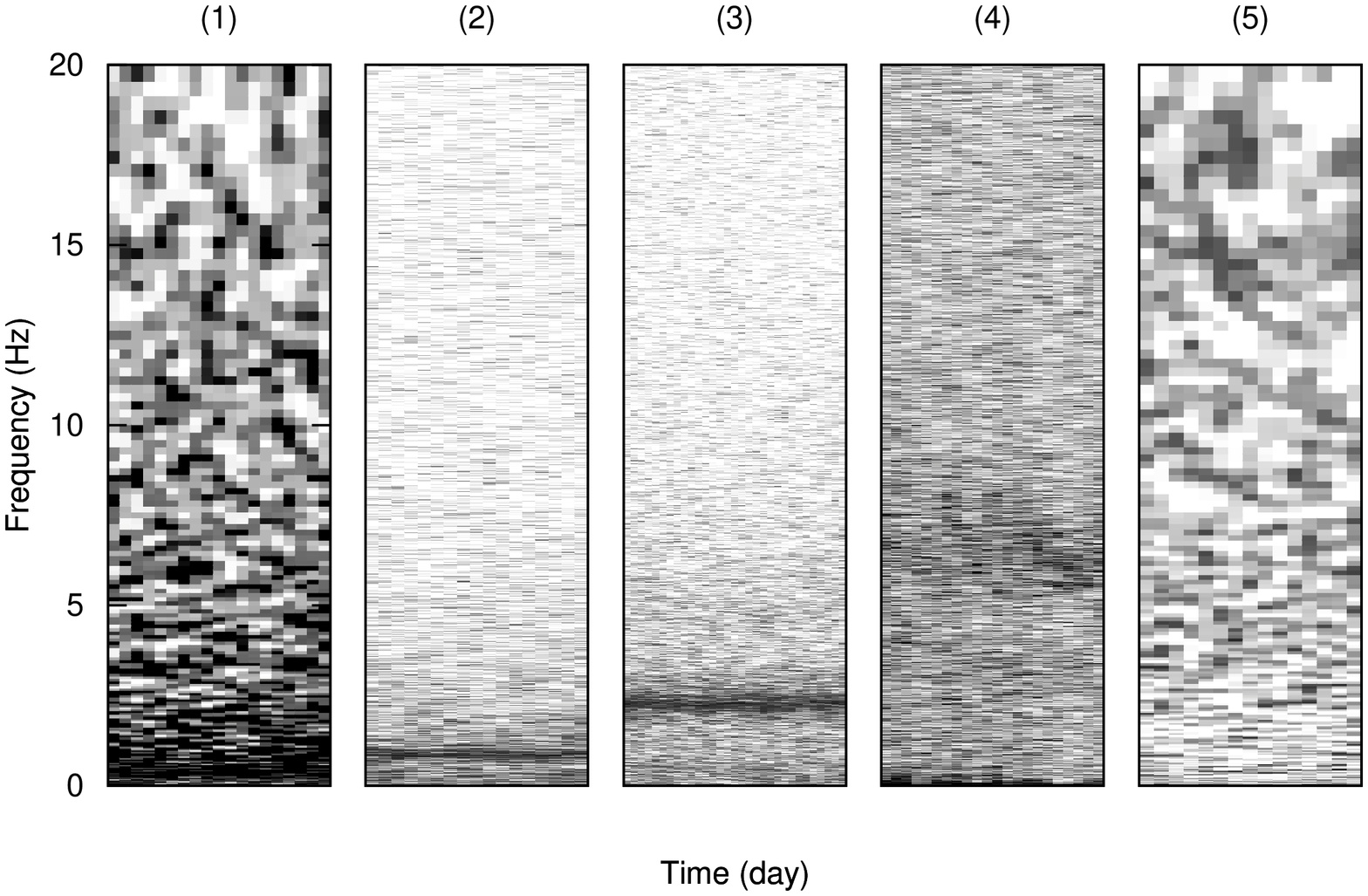}\\
\vskip -1.2 cm
\includegraphics[width=6.5cm,angle=270]{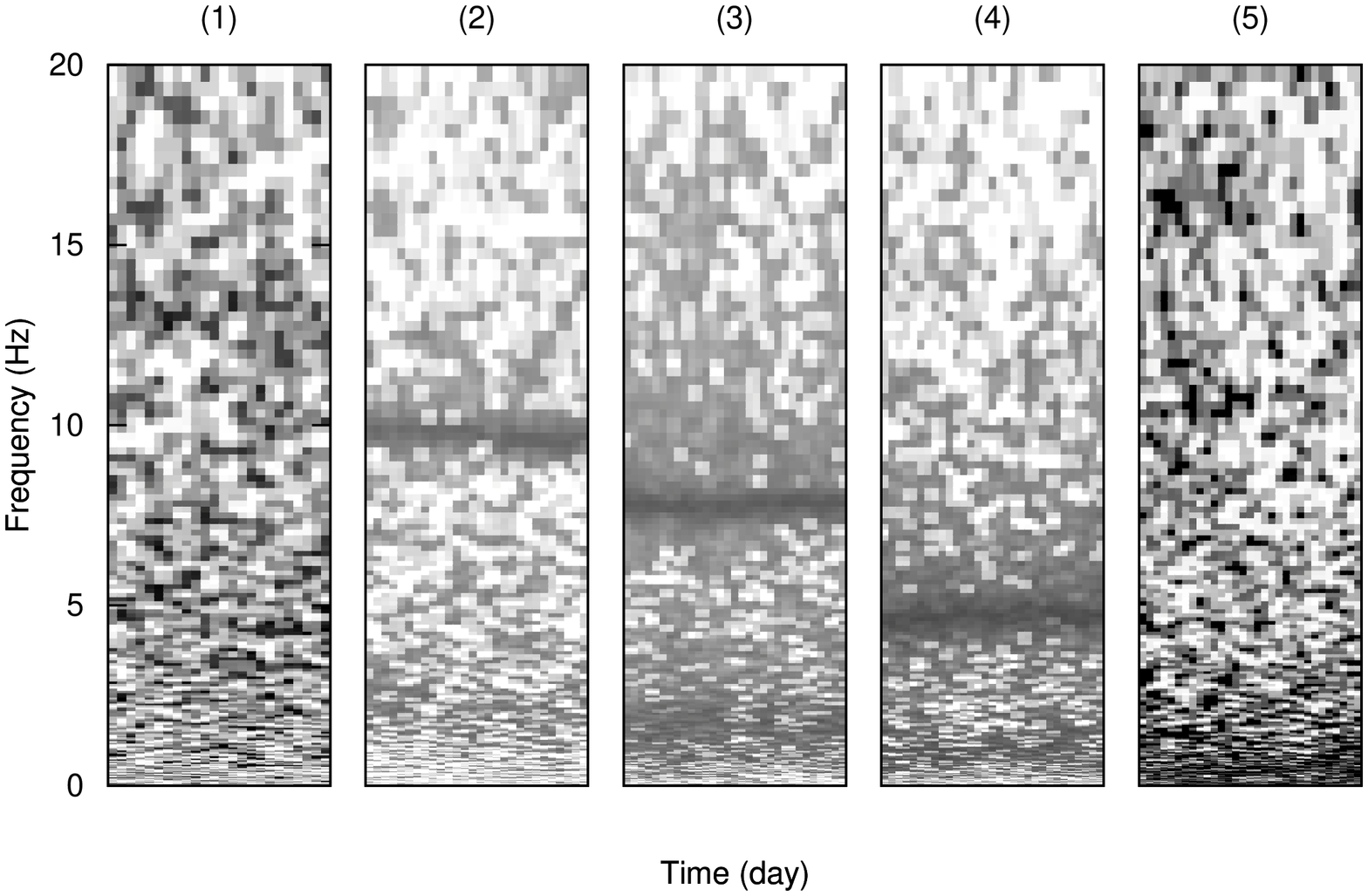}}
\vskip -0.8 cm
\caption {(a) Dynamic power density spectra over five
days in the rising phase. 
(1) Obs. ID=91404-01-01-01, QPO=0.382 Hz,
(2) Obs. ID=91702-01-01-03, QPO=0.886 Hz,
(3) Obs. ID=90704-04-01-00, QPO=2.3130 Hz,
(4) Obs. ID=91702-01-02-00, QPO=3.45 \& 6.522 Hz with a break frequency
at 0.78 Hz and (5) Obs. ID=91702-01-02-01, QPO=14.54 \& 17.78 Hz.
(b) Dynamic power density spectra over five days in the decline phase. (1) Obs. ID=91702-01-76-00, 
QPO=13.14 Hz, (2) Obs. ID=91702-01-79-01, QPO=9.77 Hz,
(3) Obs. ID=91702-01-80-00, QPO=7.823 \& 15.2 Hz with a break
frequency at 1.32 Hz, (4) Obs. ID=91702-01-80-01, QPO=4.732 Hz 
with a break frequency=0.86 Hz, (5) Obs. ID=91702-01-82-00, QPO=0.423 Hz.}
\end {figure}

\begin {figure}[b]
\vskip -1.0 cm
\centering{
\includegraphics[width=8cm]{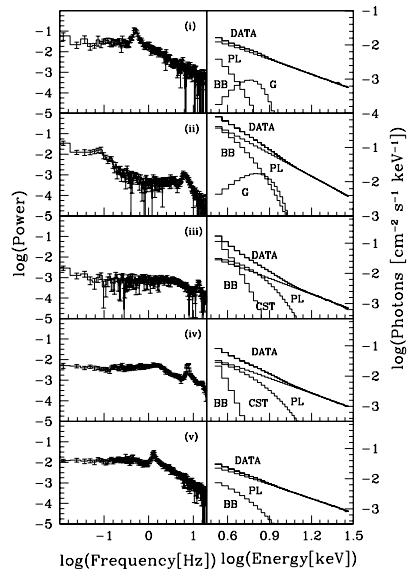}}
\caption{Power density spectrum (left panel) and energy spectral index (right panel)
of the rising phase (i and ii) and the declining phase (iii-v). 
The QPO frequencies are (i)  $0.513$ Hz (0.632), (ii) $3.45$ \& $6.522$ Hz (0.441), (iii) $13.14$ Hz 
(0.848), (iv)  $7.823$ Hz (1.14) and (v) $1.347$ Hz (1.29) respectively. Numbers next to frequencies 
are the reduced $\chi^2$ values. 
The components used for fitting the energy spectra are marked: `DATA' for total fit, 
`BB' for black body, `G' for Gaussian, `CST' for Comptonization using the Sunyaev-Titarchuk (1980) model 
with an exponential cutoff, `PL' for a power-law component without a cut-off.}
\end {figure}

In order to examine how the disk was re-adjusting itself in these two phases, 
we studied both the timing and the spectral properties  and plot the results in
Fig. 3 (panels i to v) we give an idea of this by providing
the PDS (left panels) and the fitted energy spectrum (right panels) of
five observations. The observation IDs are: 91702-01-01-01 ($11.0315^{th}$ day) and 91702-01-02-00G 
($14.6957^{th}$ day) of the rising phase and 91702-01-76-00 ($0^{th}$ day), 91702-01-80-00 
($3.27^{th}$ day) and 91702-01-81-01 ($6.118^{th}$ day) of the declining phase respectively. 
In the rising phase, the spectrum clearly becomes softer as the shock moves in and the QPO frequency increases. 
The softening of the spectrum with the increase in the QPO frequency 
has been reported by various authors (e.g., Chakrabarti et al. 2005;
ST06; Shaposhnikov et al., 2007). From a theoretical point of view,
MSC96 explicitly showed that increased cooling reduces the shock location 
and increases the QPO frequency. This was also reported by ST06
using the data of Cyg X-1. { We observe that the black body (BB) component from the
Keplerian disk and the Gaussian (G) components are also strengthened. The Gaussian component
could be from iron lines, but due to poor resolution of RXTE this cannot be said with certainty. 
In the declining phase, as the QPO frequency 
decreases, all three of the black body (BB), the power-law (PL) component
and an additional Comptonization component (using Comptonization through the Sunyaev-Titarchuk 
or CST model) from the region decreases. Interestingly, this additional cooler CST component with 
a cut-off was required only before the discontinuity observed on $3.5$th day in the 
outgoing phase, signifying that perhaps there were two sources of X-rays: one (PL) from
the post-shock region and the other (CST) from the outflow region.} Ultimately, after the
discontinuity, only a weak power-law component remains which is emitted from
a hot tenuous sub-Keplerian flow. This is all that remains after 
the outburst is over. We interpret this observation to be associated 
with a possible change in the flow behavior at $r\sim 59$: either 
the shock is propagating from the disk region to the outflow
region or the shock is moving away from the black hole due to the low 
pressure in the emptied disk. We also note that the disk component 
which was becoming stronger in the rising phase is absent in the 
late declining phase, indicating that the Keplerian component disappears 
soon after the outburst is over. 

\section{Discussions and concluding remarks}

In this Letter, we show that during the rising phase of the outburst of GRO J1655-40, 
the QPO frequency increases very slowly in the first few days and then rapidly 
increases to about $18$Hz before it disappears altogether. Our slowly drifting shock oscillation solution
explains this variation very accurately. Our estimation suggests that the shock was at 
$1270$ Schwarzschild radii when the first observation of QPO was made and it went to 
$r=59$ Schwarzschild radii when it was last observed. Within $15$ hours of this,
the shock front went inside the horizon and thus when the observation was made on the
next day, the QPO was absent. The inward drift velocity of the shock was slow, only about $20$
m s$^{-1}$. As the shock proceeds close to the black hole,
the Keplerian disk follows the shock and the energy spectrum gradually became 
softer with the black body component becoming stronger each day. Thus the whole scenario is 
consistent with our theoretical understanding (MSC96) that this drift 
is due to the reduction of the post-shock pressure by the increased cooling
effects in addition to the higher upstream ram pressure. Both MSC96 and CAM04 computed the cooling timescale
using bremsstrahlung and Comptonization respectively and showed that oscillations occur
when the infall time scale is comparable to these coolings in super-massive and
staller mass black holes respectively. As a consistency check, one could also use the 
fitted spectrum to compute the electron temperature and calculate the cooling time scale
from $E_{th}/{\dot E}$ where, $E_{th}$ is the thermal energy content of the electrons
and ${\dot E}$ is the rate of cooling due to Comptonization. However, $E_{th}$ depends on the 
accretion rate and electron temperature and ${\dot E}$ depends on the enhancement factor (Dermar
et al. 1991). However, the PCA data from 3-25keV does not allow one to compute the these unknowns
unambiguously.

In the decline phase, the QPO frequency was found to be decreasing 
monotonically. This means that an oscillating shock is propagating outwards. The nature
of the reduction of QPO frequency is very curious. For the first $\sim 3.5$ days, the shock 
location did not change very much, from $\sim 40$ to $\sim 59$ or so, as if it was stalling. 
In this phase an additional cooler Comptonized component was required.
After this, the shock suddenly moved away with almost constant acceleration for another
$16$ days before the QPO disappeared completely. The spectrum remained 
hard and the intensity decreased monotonically. Towards the end, the spectrum
is dominated only by the power-law component. We conjecture that in the first $\sim 3.5$ days the
shock remained inside the disk drifting slowly outward, perhaps due to interaction of the receding shock
with the still incoming Keplerian flow. After that, the receding Keplerian disk 
created a vacuum in the system and accelerated the shock wave
outward. This is indicated by almost square-law behavior of the shock location. 

In the two component advective flow model that we are using, the shock forms in the
sub-Keplerian component since a Keplerian flow is subsonic and cannot have a shock. At the
onset of the rising phase of the outburst, only the sub-Keplerian flow rate increases rapidly
since it is almost freely falling. The shocks we consider
may have actually formed farther out (than $1270r_g$, for example)
we could detect them only when the rms value of the QPO is high enough. Similarly, when the 
shock comes closer to the black hole horizon, the noise also rises and QPOs may not be detectable
as well.

The observational result we described here is unique in the sense that we are able to connect the 
QPO frequency of one observation with that of the next by a simple analytical formalism. To our knowledge
no competing model exists which uses a true solution of the flow, such as shocks in our case,
to explain such a behavior. In any case, even if the QPOs were generated by certain non-axisymmetric
feature (such as a blob), it is impossible that it will survive beyond a few 
orbital time-scales. Furthermore, the smooth decrease of QPO frequency requires that the blob moves outwards
through a differentially rotating disk for some weeks, odds of which is insignificant.
If our explanation of the propagation of shocks is correct, then 
this outburst shows convincingly how a shock wave smoothly disappears behind the 
horizon after $\sim 16$ days of its initial detection. Thus, such an observation 
would be unlikely in neutron star candidates. Our interpretation is generic and other 
outburst sources should also show such systematic drifts. A discussion of other similar sources will be dealt 
with in future.

\section*{Acknowledgments}

D. Debnath acknowledges the support of a CSIR scholarship 
and P.S. Pal acknowledges the support of an ISRO RESPOND project. 

{}

\end{document}